\documentclass[aps,prb,showpacs,superscriptaddress,amsmath,twocolumn]{revtex4}

\usepackage{amsmath}
\usepackage{amssymb}
\usepackage{graphicx}

\newcommand{\ContourFig}{3b\;}
\newcommand{\Hamiltonian}{(1)\;}

\newcommand{\bite}{Bi$_2$Te$_3$}
\newcommand{\bise}{Bi$_2$Se$_3$}

\begin{document}

\title{Electronic 
standing waves on the surface of the topological insulator Bi$_2$Te$_3$}

\author{P. Rakyta}
\affiliation{Department of Physics of Complex Systems,
E{\"o}tv{\"o}s University,
H-1117 Budapest, P\'azm\'any P{\'e}ter s{\'e}t\'any 1/A, Hungary
}

\author{A. P\'alyi}
\affiliation{Department of Materials Physics,
E{\"o}tv{\"o}s University,
H-1117 Budapest, P\'azm\'any P{\'e}ter s{\'e}t\'any 1/A, Hungary
}

\author{J. Cserti}
\affiliation{Department of Physics of Complex Systems,
E{\"o}tv{\"o}s University,
H-1117 Budapest, P\'azm\'any P{\'e}ter s{\'e}t\'any 1/A, Hungary
}

\begin{abstract}
A line defect on a metallic surface induces standing waves in the electronic
local density of states (LDOS).
Asymptotically far from the defect,
the wave number of the LDOS oscillations at the 
Fermi energy is usually equal to the
distance between nesting segments of the Fermi contour,
and the envelope of the LDOS oscillations shows a power-law decay as moving
away from the line defect.
Here, we theoretically analyze the LDOS oscillations close to
a line defect on the surface of the topological insulator \bite , 
and identify an important preasymptotic 
contribution with wave-number and decay characteristics markedly different
from the asymptotic contributions.
The calculated energy dependence of the wave number
of the preasymptotic LDOS oscillations is 
in quantitative agreement with the result of a
recent scanning tunneling microscopy experiment
[Phys. Rev. Lett. {\bf 104}, 016401 (2010)].
\end{abstract}



\maketitle

\section{Introduction}

Distinct surface-electronic properties,
potentially relevant for spintronic applications,
arise from the strong spin-orbit interaction in
three-dimensional topological insulators (3DTIs) \cite{Hasan-review}. 
Although the bulk electronic structure of these materials resembles that of standard
band insulators with electronic bands separated by an energy gap, 
the valence and conduction bands of the surface states form a 
conical dispersion
and touch at the center of the surface Brillouin zone.
These gapless surface states lack the standard twofold spin degeneracy,
they are protected against backscattering,
and the spin orientation of each plane-wave surface state is determined 
unambiguously by its momentum vector.

In the past few years, surface-sensitive experimental techniques
have been utilized to explore the remarkable properties of the surface electrons
in 3DTIs.
The linear, Dirac-cone-like electronic dispersion and deviations from
that were observed in various 3DTI materials 
using angle-resolved photoemission spectroscopy \cite{Hsieh-nature,Hsieh-science,
Xia-naturephysics,Chen-Bi2Te3,BiSexp,Hsieh-prl} (ARPES), and the 
correlation between spin and momentum was demonstrated 
by the spin-resolved version of the same technique \cite{Hsieh-science}.
The role of electron scattering off pointlike impurities and
line defects on 3DTI surfaces, 
highly relevant for future attempts to design electronic devices based
on these materials,
has been studied via scanning tunneling microscopy (STM)
\cite{Roushan-nature,
TongZhang-edgeimpurity,BiTexp,JungpilSeo-antimony,
JingWang}.
In the vicinity of obstacles on the surface, characteristic standing wave patterns are 
formed
due to the interference of initial and final scattering states \cite{2DEG}.
These electronic standing waves
contribute to the local density of states (local DOS, LDOS),
therefore real-space mapping of them is possible via STM.
Theories describing the standing waves on 3DTI surfaces have
also been formulated recently \cite{FuHam,WeiChengLee-prb-bite,LDOS_Zhou,Guo-ftstm,QiangHuaWang-ldos,LDOS_Biswas,Biswas_arxiv,Liu-arxiv}.

A line defect has translational symmetry in the direction it stretches along,
hence the electronic standing waves in its vicinity are essentially 
one-dimensional (1D), 
(i.e., the LDOS varies only along the axis perpendicular to the line defect).    
This simple 1D character of the induced LDOS pattern implies a relatively
straightforward experimental and theoretical analysis of the effect,
which serves as a strong motivation to consider such arrangements.
A line defect arises naturally at the edge of a step formed by
an extra crystal layer on the surface, \cite{2DEG,BiTexp,
TongZhang-edgeimpurity,JingWang} hence this 1D setup is
accessible experimentally. 

Information on the electronic system can be extracted from the asymptotic
decay exponent and wave number of LDOS oscillation around line defects. 
Theoretical results \cite{LDOS_Zhou,LDOS_Biswas,Biswas_arxiv,
JingWang} indicate that 
the LDOS oscillations on the surface of 
a 3DTI, asymptotically far from a line defect and within the energy range
of linear dispersion,
decay with the distance $x$ from the defect 
as $x^{-3/2}$.
This decay exponent is in contrast with the $\sim x^{-1/2}$ decay seen in a
standard two-dimensional electron gas, \cite{2DEG} and arises as a consequence
of the absence of backscattering characteristic of surface electrons in 3DTIs.
Recent STM data agrees with this prediction. \cite{JingWang}
The wave vector of the asymptotic LDOS oscillations is usually equal to the 
distance between nesting segments of the constant-energy contour (CEC),
which is the diameter of the Fermi circle in the above-mentioned case.
This has been used in a recent experiment \cite{TongZhang-edgeimpurity}
to confirm the linear dispersion  and to infer the Fermi velocity
on the Dirac cone in \bite.

For energies well above the Dirac point, the topological 
surface conduction band of \bite\  is subject to strong hexagonal warping. 
STM data corresponding to this energy range is available \cite{BiTexp,JingWang}, 
however, the rather complex geometry of the dispersion relation 
has so far prevented an unambiguous theoretical interpretation of the observations. 
In this work, we provide a theoretical investigation of LDOS oscillations created by
a line defect on the surface of \bite.
We describe the effect in an exact scattering-theory framework \cite{LDOS_Zhou},
yielding results that are not restricted to the spatial region asymptotically far from the
defect, but hold also in the vicinity thereof.
This enables us to directly compare our results with experimental data,
the latter being usually taken close to the defect where features
of the LDOS are most pronounced.
In the energy range of strong hexagonal warping,
we identify a significant pre-asymptotic contribution to the LDOS oscillations,
with wave number quantitatively matching that of a recent
experiment \cite{BiTexp}.

 
\begin{figure}
 \includegraphics[width=0.48\textwidth]{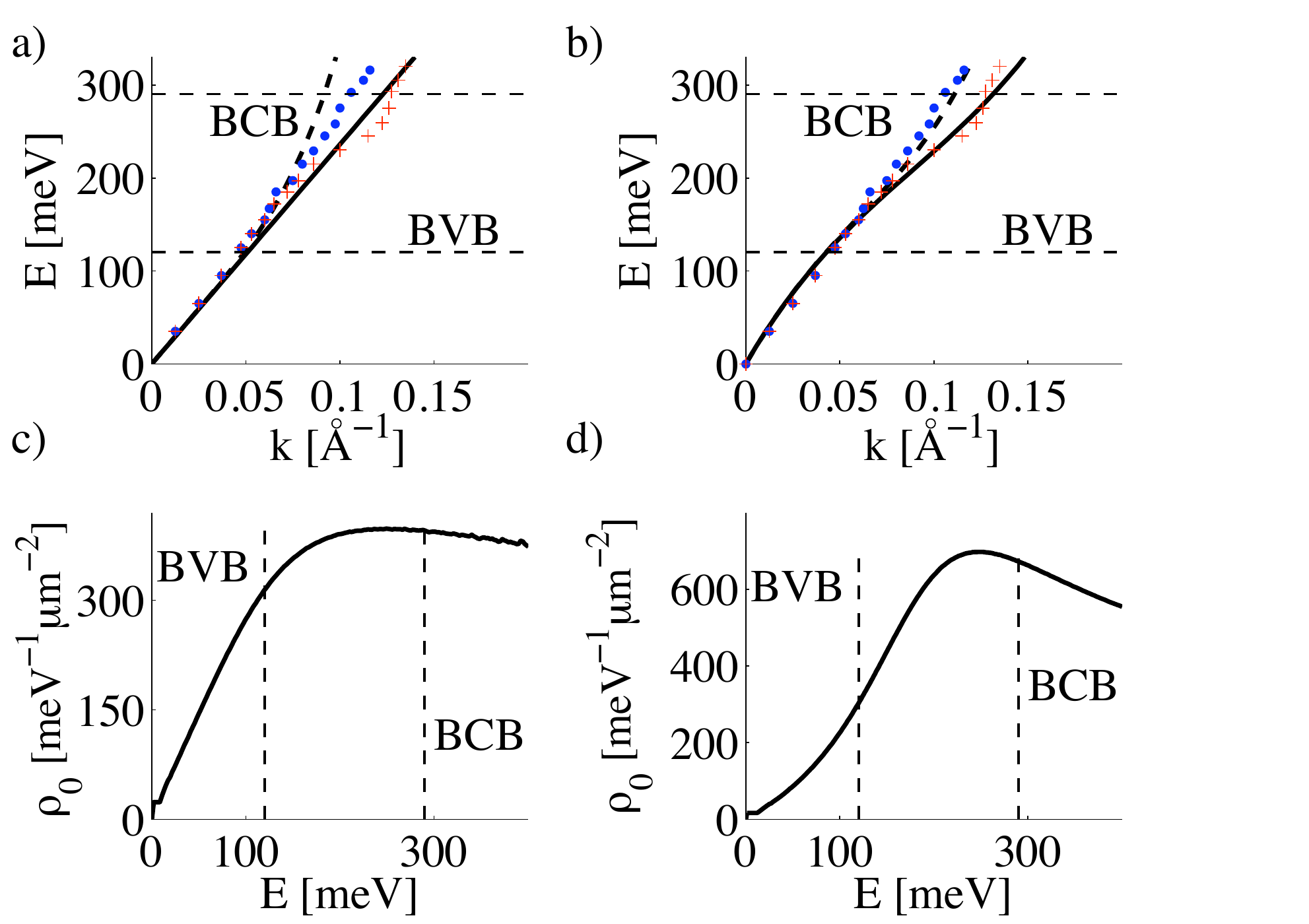}
  \caption{
  (a,b) Surface-state conduction-band dispersion of \bite \ 
  along the $\Gamma M$ (solid line) and $\Gamma K$ (dashed line)
  directions of the surface Brillouin zone. 
  (c,d)
  Surface-state conduction-band density of states of \bite .
  For (a) and (c), Eq. \eqref{eq:SSB} was used with parameter 
  values\cite{FuHam} 
  $v_0 = 2.55$ eV\AA, $\lambda = 250$ eV\AA$^3$ and
  $\alpha, \gamma =0$.
  For (b) and (d), the parameter values 
  $v_0 = 3.5\;{\rm eV\AA}$, $\lambda=150\;{\rm eV\AA^{3}}$, 
  $\alpha=21\;{\rm \AA^{2}}$ and $\gamma=-19.5\;{\rm eV \AA^{2}}$
  were used.
  Red crosses (blue points) represent the measured dispersion
  along $\Gamma M$ ($\Gamma K$) 
  (data taken from Ref.~\onlinecite{BiTexp}). 
  The zero of energy corresponds to the Dirac point of the spectrum. 
  Energy intervals overlapping with the 
  bulk valence band (BVB) and bulk conduction band (BCB) are also
  shown.} \label{fig:DOS}
\end{figure}


\section{Band-structure parameters.}
\label{sec:parameters}
In order to base our forthcoming calculations on an 
accurate surface-band dispersion, 
we first establish accurate values of the relevant 
band-structure parameters (defined below)
of \bite. 
ARPES measurements \cite{BiTexp} indicate that 
the surface bands
of 3DTIs with the crystal structure of 
\bite  \ are subject to hexagonal warping, 
which can be described by the envelope-function Hamiltonian \cite{FuHam}:
\begin{equation}
H (\mathbf{k})= 
\gamma \mathbf{k}^2 + v_{\mathbf{k}}\left(k_x\sigma_y - k_y\sigma_x \right) + 
\frac{i\lambda}{2}\left(k_+^3 - k_-^3 \right) \sigma_z\;, \label{eq:FuH}
\end{equation}
where 
$v_{\mathbf{k}} = v_0(1 + \alpha \mathbf{k}^2)$, and
$k_{\pm} = k_x\pm i k_y$. 
Here, $(\sigma_x,\sigma_y,\sigma_z)$ is the vector of Pauli matrices
representing spin,
$v_0, \lambda, \gamma$ and $\alpha$ are band-structure parameters,
$\mathbf{k} = (k_x,k_y)$,
and $k_x$ and $k_y$ are momentum components along the
$\Gamma M$ and $\Gamma K$ directions of the surface Brillouin zone, 
respectively.
For convenience, we performed a $+\pi/2$ rotation around the $z$ axis  
compared to the Hamiltonian in Ref.~\onlinecite{FuHam}.
Energy eigenvalues of $H$ in Eq. \eqref{eq:FuH} are
\begin{equation}
 \varepsilon_{\pm}(\mathbf{k}) = \gamma \mathbf{k}^2 \pm 
 \sqrt{\left(v_{\mathbf{k}} \mathbf{k} \right)^2 + \lambda^2 k_y^2 (k_y^2 - 3k_x^2)^2 }\;,\label{eq:SSB}
\end{equation}
where $+$ ($-$) stands for conduction (valence) band.
Higher-order terms\cite{Basak-prb-spintexture}
in $\bf k$ can also be included in $H$ .

Satisfactory agreement between the spectrum in
Eq.~\eqref{eq:SSB} 
and the ARPES spectra of \bise \ surface states \cite{BiSexp}
can be obtained by neglecting the band-structure 
parameters $\gamma$ and $\alpha$.
In \bite \ however, 
the conduction-band dispersion measured along the $\Gamma M$ direction,
shown with red crosses in Figs.~\ref{fig:DOS}a,b, 
has a sub-linear segment, which can be theoretically
reproduced by Eq.~\eqref{eq:SSB} 
only if $\gamma$ and $\alpha$ are finite. 
We find that using the band-structure parameter set
$v_0 = 3.5\;{\rm eV\AA}$, $\lambda=150\;{\rm eV\AA^{3}}$, 
$\alpha=21\;{\rm \AA^{2}}$ and $\gamma=-19.5\;{\rm eV \AA^{2}}$,
the measured 
dispersion relations along the $\Gamma K$ and $\Gamma M$ directions
and the 
surface density of states (Fig. 1c,d of Ref.~\onlinecite{BiTexp})
are accurately described by $H$ [Eq.~\eqref{eq:FuH}] up to 335 meV 
above the Dirac point. 
We use these parameter values throughout this paper. 
Fig.~\ref{fig:DOS}a and \ref{fig:DOS}b 
(Fig.~\ref{fig:DOS}c and \ref{fig:DOS}d)
compares the theoretical surface-state dispersion (density of states)
calculated with a parameter set used in an earlier work\cite{FuHam},
and the above parameter set that we found to be optimal, respectively. 
Further considerations used to find the optimal parameter set above
are included in Appendix \ref{sec:parameters_appendix}.


\section{Model.}
\label{sec:model}
\begin{figure}
 \includegraphics[width=0.4\textwidth]{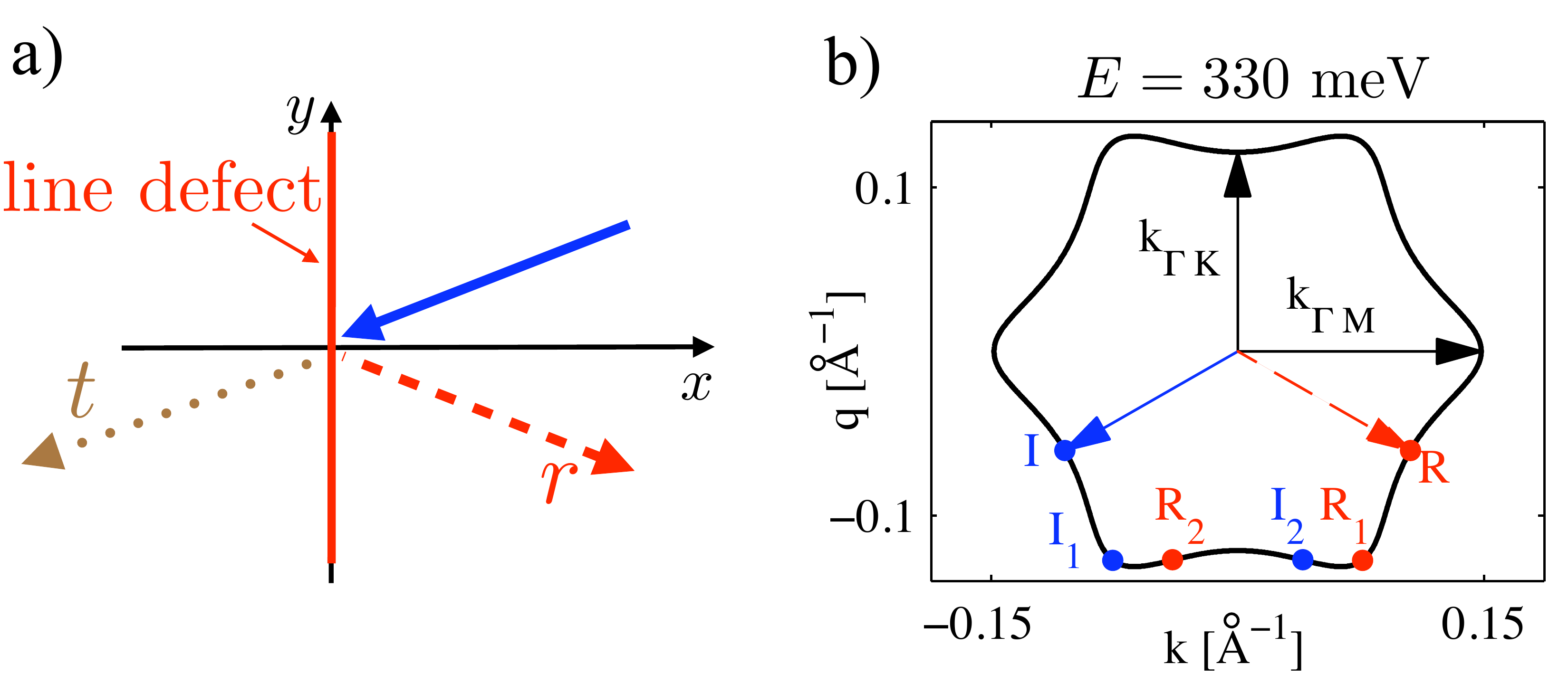}
  \caption{
   (a) An incident wave from the $x>0$ region (solid arrow)
  is partially reflected (dashed arrow) 
  and transmitted (dotted arrow) at a line defect, e.g., an atomic terrace, on the 
  surface of \bite .
  (b) Hexagonally warped constant-energy contour in reciprocal space, 
  corresponding to energy $E=330$ meV above the Dirac point.
  Incident and reflected wave vectors from (a) are also shown.}
  \label{fig:setup}
\end{figure}

Our goal is to theoretically describe the
oscillations in the surface-state conduction-band LDOS induced by a
line defect, e.g., the edge of an atomic terrace \cite{2DEG,TongZhang-edgeimpurity,
JungpilSeo-antimony,BiTexp,JingWang}, on the surface of \bite.
For the moment we assume that the defect forms a straight line that coincides with
the $y$ axis (see Fig. \ref{fig:setup}). 
Following Ref.~\onlinecite{LDOS_Zhou}, we model the system with the Hamiltonian $H+V$,
where effect of the defect is described via the potential
$V(x) = V_0 \Theta(-x)$.

Our analysis of the LDOS oscillations is based on exact
energy eigenstates describing scattering of conduction-band 
electrons by the line defect (see Fig. \ref{fig:setup}).
Therefore we first describe the scattering process of a plane wave 
energy eigenstate
$\Phi_{k,q}(x,y) e^{iqy}  = e^{iqy} e^{ikx} \chi_{k,q}$
incident
from, say, the $x > 0$ side of the edge, with
momentum components $k\equiv k_{x}$ and $q \equiv k_{y}$,
energy $E$, and spin wave function $\chi_{k,q}$.
Scattering at the line defect is elastic, hence the energy $E$ is conserved.
The momentum component $q$ parallel to the defect is 
also conserved due to translational invariance in the $y$ direction. 
The value of $q$ 
determines the number of propagating waves at a given energy. 
For example,  in Fig. \ref{fig:setup}b, the number of propagating waves can
be two (I and R) or four (I$_1$, I$_2$, R$_1$, R$_2$), depending on $q$.

However, the incident plane wave can be scattered into
coherent superpositions of three reflected and three transmitted partial waves,
for the following reasons.
On the $x>0$ side of the defect, 
the equation $E=\varepsilon_+ (k_r,q)$ has six complex solutions 
$k_{r,1},\dots k_{r,6}$ 
for given values of $E$ and $q$,
which follows from Eq.~\eqref{eq:SSB} 
(the numerical method used to obtain these solutions is 
described in Appendix \ref{sec:planewave}).
Three of the $k_{r,p}$-s correspond to propagating waves moving
away from the defect or evanescent modes.
The associated wave functions 
$\Phi_{k_{r,p},q} = e^{iqy}e^{ik_{r,p}x} \chi(k_{r,p},q)$ ($p=1,2,3$)
should be included in the Ansatz of the complete scattering state.
The remaining three solutions $k_{r,p}$ $(p=4,5,6)$ correspond
to propagating waves towards the defect or diverging modes,
hence they are disregarded. 

These arguments, together with their generalization to 
transmitted waves, imply that the $x$-dependent
component of the complete scattering wave function is
\begin{eqnarray}
\psi^{(R)}_{k,q}(x) =
  \left\{ \begin{array}{cc}
  \Phi_{k,q}(x) + 
  \sum\limits_{p=1}^3 
     r_{kq,p} \alpha_{kq,p} \Phi_{k_{r,p},q}(x)  &  \textrm{if } x > 0\;, \\
  \sum\limits_{p=1}^3 t_{kq,p} \beta_{kq,p} \Phi_{k_{t,p},q} (x)  & \textrm{if } x < 0 \;, \label{eq:Psi}
  \end{array} \right. 
\end{eqnarray}
where
\begin{eqnarray}
 \alpha_{kq,p} &=& 
   \left\{
   	\begin{array}{ll}
		\sqrt{\frac{|v_{\perp,k,q}|}{|v_{\perp,k_{r,p},q}|}} & 
		\textrm{if } k_{r,p} \in \mathbb{R}, \\
		1 & \textrm{otherwise},     
	\end{array}
   \right.      \\
 \beta_{kq,p}  
 &=& 
   \left\{
   	\begin{array}{ll}
		\sqrt{\frac{|v_{\perp,k,q}|}{|v_{\perp,k_{t,p},q}|}} & 
		\textrm{if } k_{t,p} \in \mathbb{R}, \\
		1 & \textrm{otherwise}.
	\end{array}
   \right. 
\end{eqnarray}
Here, the $r$-s and $t$-s are reflection and transmission coefficients, 
$v_{\perp,k,q}$ is the $x$-component of the group velocity of
the plane wave with wave-vector components $(k,q)$,
and the factors $\alpha$ and $\beta$ ensure the unitarity of the scattering
matrix built up from the reflection and transmission coefficients.
Evanescent modes are not subject to the unitarity requirement, hence
we are allowed to make the above arbitrary choice 
$\alpha = \beta = 1$ for partial waves with complex wave vectors.
As the Hamiltonian is a third-order differential operator, 
partial waves at the two sides of the defect should be matched via
boundary conditions ensuring their continuity as well as the continuity
of their first and second derivatives.
The scattering state $\psi_{k,q}^{(L)}$ of a plane wave incident
from the left side ($x<0$) of the defect can be described analogously.

\begin{figure*}
 \includegraphics[width=1\textwidth]{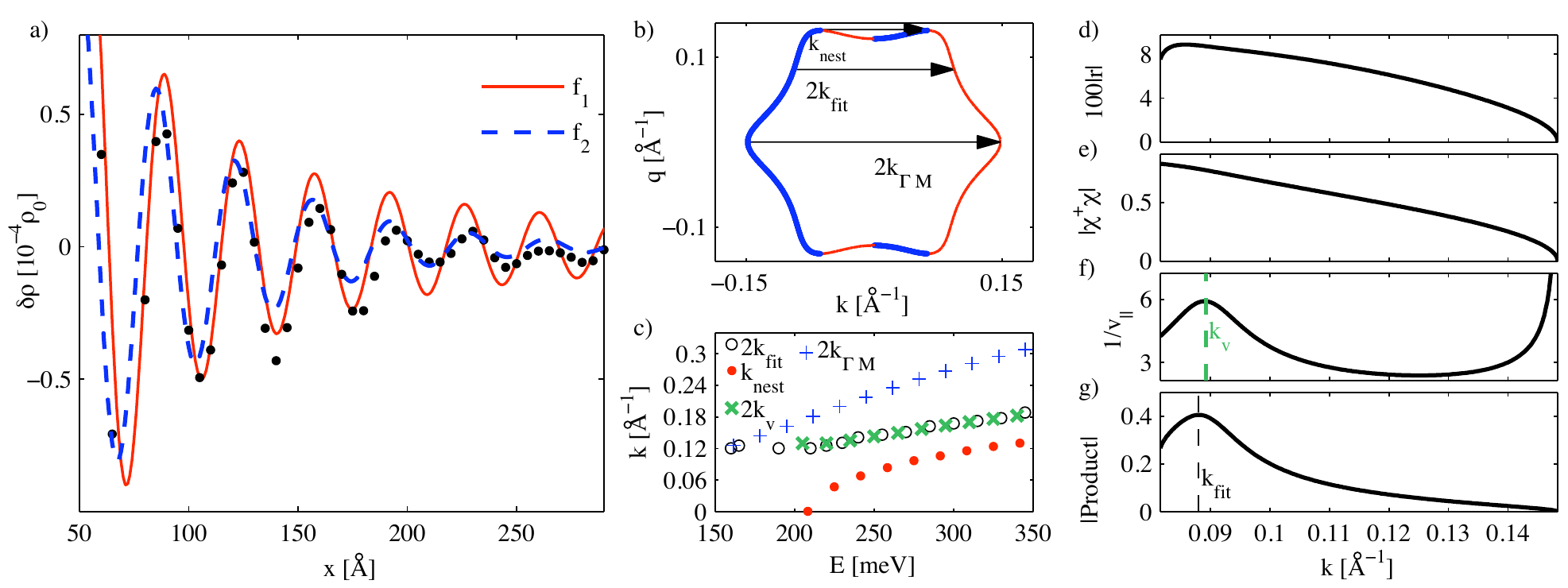}
  \caption{
  (a) Position-dependent contribution $\delta \rho(E,x)$ (solid line)
  to the LDOS at energy
  $E = 330$ meV in the vicinity of a line defect (black points). 
  Red solid and blue dashed lines are 
  fits of the functions $f_1$ and $f_2$ (see text), respectively, 
  to the data.
  $\rho_0 = 628 \ {\rm meV}^{-1} \mu {\rm m}^{-2}$.
  (b) Constant-energy contour at the same energy,
  and relevant scattering wave vectors in
  reciprocal space.
  Thick blue (thin red) pieces of the contour correspond to
  left-moving (right-moving) plane waves.
  The union of the thick blue pieces form $\Gamma_{E}^{(R)}$
  in Eq. \eqref{eq:contour}.
  (c) Dominant wave number $2k_{\rm fit}$ of $\delta \rho(E,x)$ (open circles),
  and characteristic wave numbers 
  $k_{\rm nest}$ (red points),
  $k_{\rm v}$ (green diagonal crosses)
  and $2k_{\Gamma {\rm M}}$ (blue crosses) of the constant-energy contour,
  as functions of energy $E$.
  (d) Magnitude of reflection amplitude $|r| \equiv |r_{k,q'}|$,
  (e) magnitude of spinor overlap 
  $|\chi^\dag \chi |\equiv |\chi_{k,q'}^\dag \chi_{-k,q'}|$,
  (f) magnitude of the inverse of the parallel-to-defect
  group velocity component $v_\parallel(k)$ 
  (in units of $10^{-6}\, {\rm s}/{\rm m}$),
  and
  (g) the product of the above three quantitites 
  (in units of $10^{-6}\, {\rm s}/{\rm m}$),
  as functions of perpendicular-to-defect wave number component $k$.
  } \label{fig:results}
\end{figure*}

The LDOS at a given energy $E$ and position $x$ is
expressed with the exact scattering states as 
\begin{equation}
\label{eq:contour}
\rho(E,x) = 
\frac{1}{(2\pi)^2 \hbar} \sum_{d=L,R}
\int_{\Gamma_E^{(d)}} d\kappa
\frac{|\psi_{k,q}^{(d)}(x)|^2}{v(k,q)},
\end{equation}
where 
$\Gamma_E^{(R)}$ [$\Gamma_E^{(L)}$] is the wave-vector
contour of waves that 
(i) are incident from the right [left] side of the line defect, and
(ii) have energy $E$.
Note that $\Gamma_E^{(d)}$ breaks up to disconnected pieces 
for strong hexagonal warping,
e.g., in Fig. \ref{fig:setup}b, the points $I_1$ and $I_2$ belong to $\Gamma_E^{(R)}$
but $R_1$ and $R_2$ do not.
(The wave-vector contour $\Gamma_E^{(R)}$ is shown in 
Fig. \ref{fig:results}b,
there it is formed as the union of the thick blue lines.)
The infinitesimal line segment along
$\Gamma_E^{(L,R)}$ is denoted by $d\kappa$, 
and $v(k,q)$ is magnitude of the group velocity of the wave
with momentum vector $(k,q)$.
Using the exact scattering wave functions $\psi^{(d)}_{k,q}$,
we evaluate the integral in Eq. \eqref{eq:contour} numerically.
To account for the inevitable roughness of the line defect and to
suppress noise due to the limited precision of the
numerical integration, we average the LDOS oscillations $\delta \rho(x)$
over the angular range $[-5^\circ,5^\circ]$ of the line defect orientation with
respect to the $\Gamma K$ direction.


\section{Results.}
\label{sec:results}
In Fig. \ref{fig:results}a, 
we plot the numerically computed LDOS oscillations
$\delta \rho (x) \equiv \rho(x)-\rho_0$ (black points)
on the $x>0$ side 
of 
the defect, corresponding to 
energy $E=330$ meV and potential step height $V_0 = -150$ meV. 
Recent theories using asymptotic analysis \cite{JingWang,Biswas_arxiv}
suggest that the LDOS oscillations in the vicinity of a line
defect on the surface of \bite \ decay 
no faster than $x^{-3/2}$.
Motivated by this finding, 
we fit the function $f_1(x) = A_1 \sin (2k_1 x + \varphi_1) x^{-3/2}$ 
via fitting parameters $A_1$, $k_1$, and 
$\varphi_1$ to the results (shown as red solid line).
We also fit an exponentially decaying function 
$f_2(x) = A_2 \sin (2k_{\rm fit} x+ \varphi_2) e^{-x/L}$ via fitting 
parameters $A_2$, $k_{\rm fit}$, $\varphi_2$ and $L$ (blue dashed line).
The two major features of our numerical result $\delta \rho(x)$ are as follows. 
(i) 
Comparison of the three curves suggests that
the decay of $\delta \rho(x)$ is better described by the exponentially decaying 
function $f_2(x)$ than by $f_1(x)$ having power-law decay
(see Appendix \ref{sec:fourier} for further details).
(ii)
The wave-number value obtained from fitting $f_2(x)$ is 
$2k_{\rm fit} \approx 0.178$ \AA$^{-1}$.

Expectations for the wave number of the 
LDOS oscillations can be drawn from 
asymptotic analysis \cite{JingWang,Biswas_arxiv}.
That suggests that the wave number of an
electronic standing wave at a given energy $E$ is associated to wave vectors
connecting nesting segments of the corresponding CEC \cite{2DEG,FuHam,Biswas_arxiv},
i.e., $2 k_{\Gamma M} = 0.297$ \AA$^{-1}$
or $k_{\rm nest} = 0.126$ \AA$^{-1}$ depicted in Fig.~\ref{fig:results}b.
As the wave number $2k_{\rm fit}$ characteristic of our data $\delta\rho(x)$
deviates significantly from $2 k_{\Gamma M}$ and 
$k_{\rm nest}$, and its decay is exponential rather than power-law,
we conclude that $\delta\rho(x)$ is dominated by a pre-asymptotic contribution
in the considered spatial range.

In what follows, we argue that (i) the pre-asymptotic component of the 
LDOS is due to the interference of incoming and reflected partial waves, i.e.,
the role of evanescent and transmitted partial waves is negligible, and
(ii) the appearance of the characteristic wave number $2k_{\rm fit}$ in the 
LDOS oscillations is due to the non-monotonic behavior of the 
parallel-to-defect group velocity component along the CEC.
To this end, we now consider only the interference contribution
of incoming and reflected propagating waves
to the LDOS [Eq. \eqref{eq:contour}] on the right half plane $\rho(E,x>0)$,
rewrite it as an integral over the perpendicular-to-defect wave-vector component
$k$, and drop the contributions from $k$-regions where more than 
one propagating reflected partial wave is allowed ($|k|<k_c$),
yielding
\begin{equation}
\label{eq:rhor}
\rho_r(E,x) = 
\frac{1}{2 \pi^2 \hbar} 
\int_{-k_{\Gamma M}}^{-k_c} dk
\frac{\left(r_{k,q'} \chi^\dag_{k,q'} \chi_{-k,q'} e^{-i2kx} + c.c.\right)}{|v_\parallel(E,k)|}.
\end{equation}
Here, $q' \equiv q'(k,E)$ is the unique positive solution of $E=\varepsilon_+(k,q)$
for a fixed $k$ and $E$.
The integrand without the exponential factor is related to the
Fourier transform of $\rho_r(x)$ . 
We plot the three factors determining $\rho_r(x)$ --- 
the magnitudes of the reflection coefficient $|r_{k,q'}|$,
the spinor overlap $|\chi^\dag_{k,q'} \chi_{-k,q'}|$,
and the inverse of the parallel-to-defect 
group-velocity component $v_\parallel(k,E)$, ---
as well as their product,
in Fig. \ref{fig:results}d, e, f, and g, respectively.
While Figs. \ref{fig:results}d and e show a featureless
dependence on $k$,
Fig. \ref{fig:results}f
reveals a peak in $1/v_\parallel(k)$.
The corresponding local maximum point, which we denote
with $k_v$, is very close to $k_{\rm fit}$ obtained from the numerical
result in Fig.~\ref{fig:results}a.
As a consequence of the peak in $1/v_\parallel(k)$, 
a peak arises in the product of the three factors (Fig. \ref{fig:results}g)
as well.
This analysis reveals that the characteristic wave number
 $k_{\rm fit}$ of the LDOS oscillation $\delta \rho(x)$ 
is determined, to a large extent, by the electronic dispersion relation
via $v_\parallel(k)$, and the
details of the scattering process have little significance on its value.

We have repeated the above analysis for various energy values in
the range $E\in [145\ {\rm meV}, 475\ {\rm meV}]$ 
in order to compare the characteristic wave number $2 k_{\rm fit}$
of $\delta \rho(x)$ with experimental data \cite{BiTexp},
and to confirm the correlation between 
the characteristic wave numbers obtained from the dispersion relation
[$k_v(E)$] and from the numerical LDOS calculation [$k_{\rm fit}(E)$].
We plot $2k_{\rm fit}$ as the function of energy $E$ in Fig.~\ref{fig:results}c.
For low energy $E \lesssim 170$ meV, the hexagonal warping of the CEC is weak,
and our result shows $k_{\rm fit} \approx k_{\Gamma M}$ and 
a decay of $\delta \rho (x) \propto x^{-3/2}$ (not shown in the figures), 
in agreement with the results of the asymptotic analysis \cite{JingWang,Biswas_arxiv}.
Between 190 meV and 345 meV above the Dirac point, 
our $k_{\rm fit}$ data in Fig. \ref{fig:results}c differs significantly from $k_{\Gamma M}$,  and 
the former shows good quantitative agreement
with the experimental values (shown in Fig. 4b of Ref.~\onlinecite{BiTexp}).
Remarkably, $2k_v(E)$, shown as green diagonal crosses in
Fig.~\ref{fig:results}c, is almost perfectly correlated with $2k_{\rm fit}$,
confirming the generality of the above proposition that the 
characteristic wave number of the LDOS oscillation is determined
by the electronic dispersion.

No experimental data is available below 190 meV, whereas above
345 meV, the measured data 
(shown in Fig. 4b of Ref.~\onlinecite{BiTexp})
shows a pronounced kink that is not described by our model.
A potential reason for that discrepancy is that 
the surface and bulk conduction electrons might be strongly hybridized
in that high-energy range, making our surface-band model inappropriate to describe
the corresponding standing-wave patterns.

At high energy $E > 200$ meV, the nesting of CEC segments connected
by $k_{\rm nest}$ in Fig. \ref{fig:results}b may also induce LDOS oscillations
with wave number $k_{\rm nest}$\cite{JingWang,Biswas_arxiv}.
However, in our model we find that such oscillations do not exist, due to
the exact cancellation of contributions from reflected and transmitted 
waves incident from the $x>0$ and $x<0$ regions, respectively
(see Sec.~\ref{sec:cancellation}).

The appearance of the LDOS contribution with wave number corresponding
to the local maximum point of $1/v_\parallel(k)$ is a generic feature,
expected to be present in other electronic systems as well.
We think that it plays a dominant role in \bite \
because of the suppression of the other two Fourier components 
with wave numbers $k_{\rm nest}$ and $2k_{\Gamma M}$, due
to the cancellation mechanism (see Sec.~\ref{sec:cancellation}) and
the absence of backscattering, respectively. 

\section{Irrelevance of transmitted waves to the LDOS oscillation}
\label{sec:cancellation}

In principle, plane waves incident from the $x<0$ region can 
contribute to the LDOS oscillations in the $x>0$ region, provided
that they are transmitted into at least two propagating channels
on the $x>0$ side of the line defect. 
In this section, we show that such a contribution is precisely balanced
and canceled out in our model 
by the interfering reflected components of
plane waves incident from the $x>0$ side.
In turn, this cancellation is responsible for the absence of LDOS
oscillations with wave number $k_{\rm nest}$.
Without the above cancellation mechanism, such oscillations
would be expected to arise
as $k_{\rm nest}$ connects nesting segments of the
CEC (see Fig.~\ContourFig).

As we show now, our 
above statements follow from the unitary character of the
scattering matrix describing the line defect. 
Following the notation used in Sec.~\ref{sec:model}, 
consider electron plane waves with energy $E$ and
parallel-to-defect wave vector component $q$. 
We assume for concreteness that for these given parameters 
$E$ and $q$, there exists two incoming,
and correspondingly, two outgoing plane waves on
the $x>0$ side, and one incoming and one outgoing
wave on the $x<0$ side.
The corresponding perpendicular-to-defect wave-vector components
are $k_{{\rm i}1}$, $k_{{\rm i}2}$, $k_{{\rm o}1}$, 
$k_{{\rm o}2}$ on the $x>0$ side and
$\tilde k_{\rm i}$ and $\tilde k_{\rm o}$ on the $x<0$ side, respectively.
In this example, the scattering matrix $S(E,q)$ has the
following structure\cite{Datta}:
\begin{equation}
S = \left(
\begin{array}{ccc}
r & t'_1 & t'_2 \\
t_1 & r'_{11} & r'_{12} \\
t_2 & r'_{21} & r'_{22} 
\end{array}
\right).
\end{equation}
A specific example is shown in Fig.~\ref{fig:setup}b, 
where the points $I_1\leftrightarrow k_{{\rm i}1}$ and 
$I_2 \leftrightarrow k_{{\rm i}2}$ represent the incoming waves
from the $x>0$ region and $R_1 \leftrightarrow k_{{\rm o}1}$ and 
$R_2\leftrightarrow k_{{\rm o}2}$ represent the
reflected and transmitted waves in the $x>0$ region.
Note that the conclusions of this Section hold for different number of 
scattering channels as well.

Each of the two electron waves incoming from the $x>0$ side
is reflected into two propagating states with 
reflection amplitudes $r'_{p' p}$ ($p,p'=1,2$).
The incident wave from $x<0$ 
is transmitted into two propagating states on the $x>0$ side
with transmission amplitudes $t_{p'}$ ($p'=1,2$).
Straightforward calculation shows that the
contribution of the transmitted waves to the LDOS oscillations in 
the $x>0$ region is accompanied by a contribution from the
interference of the reflected waves, and the sum of these
contributions is proportional to 
\begin{equation}
\label{eq:cancel}
\left({r'}^*_{11} r'_{21} + {r'}^*_{12} r'_{22} + t^*_1 t_2\right)
e^{i(k_{{\rm o}2}-k_{{\rm o}1})x}.
\end{equation}
The first factor in Eq.~\eqref{eq:cancel} 
is the scalar product of the first and second rows of the
scattering matrix describing the line defect, hence it vanishes 
because of the unitary character of the scattering matrix. 

This finding has the following remarkable consequence with respect to
our calculated LDOS oscillations $\delta \rho(x)$.
In the $q$ intervals where two incoming and two outgoing
waves exist (an example with a specific $q$ is shown in 
Fig.~\ref{fig:setup}b,  
$I_1\leftrightarrow k_{{\rm i}1}$, $I_2 \leftrightarrow k_{{\rm i}2}$,
$R_1 \leftrightarrow k_{{\rm o}1}$ and $R_2\leftrightarrow k_{{\rm o}2}$)
the wave number $k_{{\rm o}2} - k_{{\rm o}1}$
approaches $k_{\rm nest}$ in a stationary fashion as
$q$ approaches its extremal value on the CEC,
which, in principle, implies that the wave number  $k_{\rm nest}$
is visible in the LDOS oscillations.
In practice, however, the prefactor of the term 
oscillating with $k_{{\rm o}2} - k_{{\rm o}1}$, i.e., the first factor
in Eq.~\eqref{eq:cancel}, is zero for the whole $q$ range with multiple 
outgoing waves.

Note that the wave numbers $k_{{\rm i}p}-k_{{\rm o}p'}$ ($p,p' = 1,2$),
which appear in the LDOS oscillations due to interference between
incoming and reflected waves,
do approach $k_{\rm nest}$ as $q$ approaches its extremal 
value on the CEC, but not in a stationary fashion.
Consequently, these interference terms are also not able to 
promote $k_{\rm nest}$ to the dominant wave number of the LDOS 
oscillations. 
In summary, both the theoretical findings presented in this section
and our numerical results shown above indicate that in the considered
parameter range, the characteristic wave number of the LDOS oscillation 
in the vicinity of the line defect is not $k_{\rm nest}$.


\section{Discussion and conclusions}
A line defect on a metallic surface is usually modeled as a 
potential step \cite{JingWang,2DEG,LDOS_Zhou} 
or as a localized potential barrier \cite{LDOS_Biswas,Liu-arxiv,related} 
(e.g., Dirac-delta potential). 
In our work, we make the convenient, but arbitrary choice of
modeling the defect as a potential step. 
It is important to note that the fine details of the LDOS 
results might in fact depend on the choice of the model of
the defect (step vs. localized barrier).
However, our interpretation explaining our main result,
i.e., the dependence
of the standing wave's 
wave number $k_{\rm fit}(E)$ 
on the Fermi energy $E$, is
based on the momentum-dependence 
of the inverse velocity $1/v_{\parallel}$ (see Figs.~\ref{fig:results}c-g).
The latter quantity is independent of the model describing the line 
defect, therefore our main conclusion is expected to hold 
even if 
(a) the height of the potential step at the line defect, denoted by $V_0$, 
is varied, or 
(b) a different model (e.g., Dirac-delta potential barrier) is used to 
represent the defect. 
To further support the expectation (a), we have numerically 
calculated the LDOS oscillations for various values of the potential step 
height $V_0$ 
and found no qualitative change in the inferred $k_{\rm fit}(E)$
data.
\begin{figure}
 \includegraphics[width=0.48\textwidth]{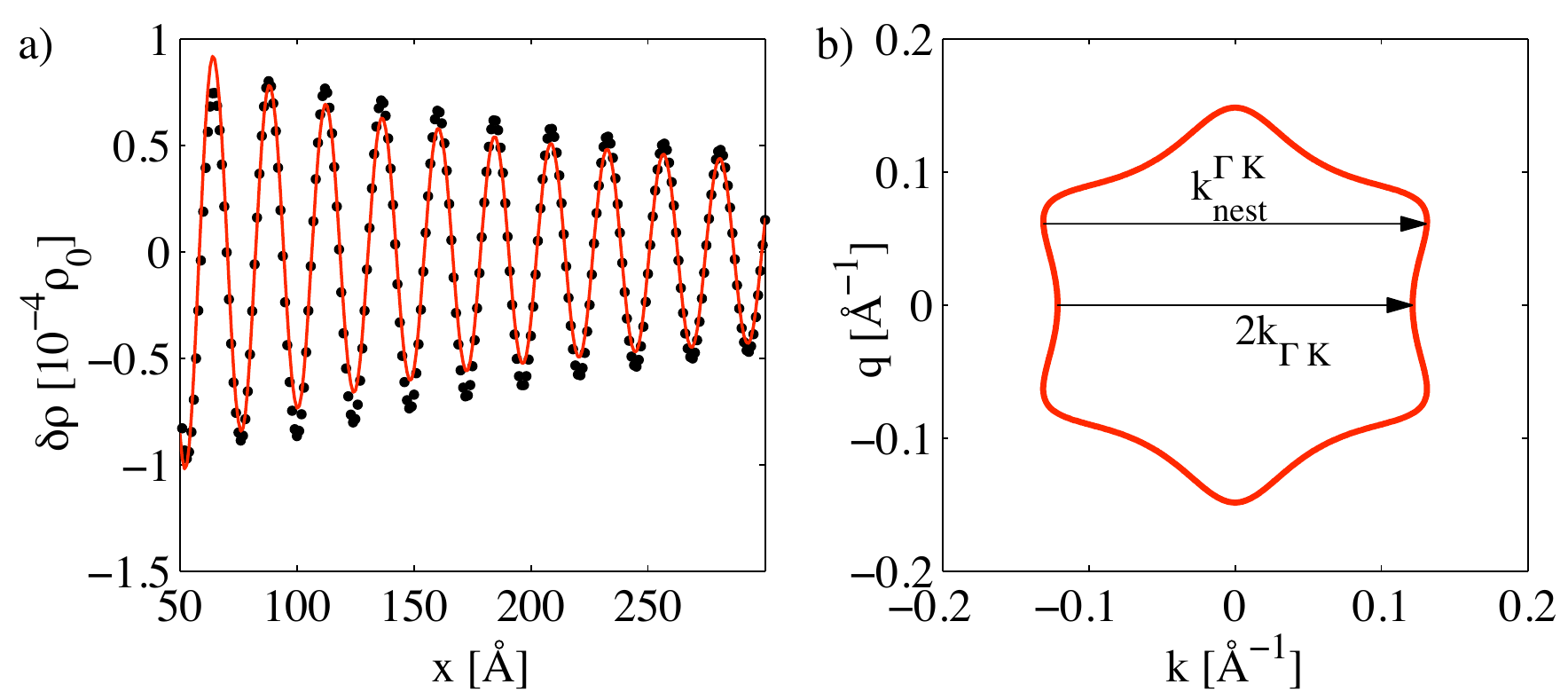}
  \caption{(a) Numerically obtained LDOS oscillations (points) 
  at energy $E=330$ meV
  for
  the case when the line defect is perpendicular to the 
  $\Gamma K$ direction,
  and the fit of $f(x) = A \sin(k x + \varphi)/x^{1/2}$ with
  fitting parameters $A$, $k$ and $\varphi$ (solid red line).
  (b) Constant energy contour and the 
  relevant wave numbers $k_{\rm nest}$ and $2k_{\Gamma K}$.
  The wave number of the oscillation in (a) is given
  by $k_{\rm nest}^{\Gamma K}$, 
  as the oscillations with $2k_{\Gamma K}$ 
  are fast decaying ($\propto x^{-3/2}$) 
  due to the absence of backscattering.}
  \label{fig:LDOS_GK}
\end{figure}

Even though the line defect in the considered experiment Ref.
\onlinecite{BiTexp}
was perpendicular to the $\Gamma M$ direction of the surface
Brillouin zone\cite{LDOS_Biswas},
it is instructive, and regarding future experiments, potentially
useful to consider the other high-symmetry case when 
the line defect is oriented perpendicular to the $\Gamma K$ 
direction. 
In Fig. \ref{fig:LDOS_GK}, we demonstrate 
that both the wave vector and the decay 
characteristics of the LDOS standing waves we obtain from
our numerical technique are in 
complete correspondence with the analytical results
of asymptotic analysis\cite{JingWang, Biswas_arxiv}.
Namely, the wave number of the oscillation is given 
by the extremal (maximal)
perpendicular-to-defect width $k_{\rm nest}^{\Gamma K}$
of the constant energy contour,
whereas the decay goes as $\delta \rho(x) \propto 1/\sqrt{x}$.
Oscillations with $2k_{\Gamma K}$ are not seen in
Fig.~\ref{fig:LDOS_GK}a presumably because they are 
fast decaying ($\propto x^{-3/2}$) 
due to the absence of backscattering\cite{LDOS_Biswas}.

In conclusion,
we theoretically described pre-asymptotic electronic LDOS oscillations 
in the vicinity of a line defect on the surface of \bite, 
with wave number and decay characteristics 
markedly different from the asymptotic ones.
The calculated energy dependence of the characteristic wave number of the
LDOS oscillations is in line with STM data.
In a general context, our study highlights the importance of 
pre-asymptotic calculation of the surface-state LDOS oscillations
in the analysis and interpretation of STM experiments.

\textit{Note:} While completing this manuscript, we became aware of a 
related  work \cite{related} on electronic standing waves on 3DTI surfaces.
Our results partially overlap with those in Ref. \onlinecite{related}.
Apart from various details of the model, 
the major distinctive  features of our work are 
(i) the interpretation of the results in
terms of the properties of the group velocity (Fig. \ref{fig:results}d-g),
and (ii) the quantitative agreement with the  experimental results
reported in Ref. \onlinecite{BiTexp}.

\acknowledgements
This work was supported by 
OTKA grants No. 75529, No. 81492, and
No. PD100373,
the Marie Curie ITN project NanoCTM,
the European Union and co-financed by the European Social Fund 
(grant no.~TAMOP 4.2.1/B-09/1/KMR-2010-0003),
and the Marie Curie grant CIG-293834.

\appendix

\section{Band-structure parameters of the homogeneous system}
\label{sec:parameters_appendix}

To base our calculation on an accurate surface-state dispersion relation,
in Sec. \ref{sec:parameters}. 
we estimated the band-structure parameters of 
\bite \ by comparing the theoretical dispersion (see Eq.~\eqref{eq:SSB})
and DOS with
the experimentally observed ARPES and STM spectra and the
DOS derived from those.
The four band-structure parameters are
$v_0$, $\lambda$, $\gamma$ and $\alpha$.
Here we outline the considerations we used for those estimates.

The signs of $\gamma$ and $\alpha$ 
can be determined by considering
 the ARPES spectrum along the $\Gamma M$ direction, 
shown as red crosses in Fig.~\ref{fig:DOS}a and b.
Note that this cut of the dispersion relation corresponds to the
function $\varepsilon_+(k_x,0)$ in Eq.~\eqref{eq:SSB}.
The measured dispersion is linear for small wave number,
its slope first becomes smaller as the wave number is increased,
but then the slope increases again as wave number is increased further. 
This characteristic is naturally captured by a polynomial of the wave 
number with negative second-order coefficient and positive 
third-order coefficient.
Since the third-order Taylor series of $\varepsilon_+(k_x,0)$
in $k_x$ around zero has the form
\begin{equation}
\varepsilon_+(k_x,0) \approx  
v_0 k_x + \gamma k_x^2 +\alpha v_0 k_x^3
\end{equation}
we can conclude that $\gamma<0$ and $\alpha>0$
is required to describe  the measured dispersion. 
The signs of the remaining two parameters $v_0$ and $\lambda$
has no effect on the spectral properties, therefore we assign
a positive sign to them. 

Having the signs of band-structure parameters established, 
we determined the values of the four parameters
by systematic visual comparison of the experimental 
dispersions (Figs.~\ref{fig:DOS}a,b),
the DOS data obtained from ARPES and STM measurements
 (Figs. 1c,d of Ref. \onlinecite{BiTexp}),
 and the corresponding theoretical curves.

\section{Plane-wave states}
\label{sec:planewave}

We label the electronic plane-wave 
states by their energy $E$ and the 
parallel-to-defect wave-number component $q$, 
which are conserved in the scattering process on the line defect 
(for details see the main text).
Here we review a numerical method to obtain these states,
which is necessary to solve the scattering problem at the edge step.
For a given $E$ and $q$ there are six solutions of longitudinal wave vector $k_{r,p}$ ($p=1\dots 6$) which satisfies the characteristic equation 
$\det\left[H(k_r,q) - \rm{\hat{I}}E\right] = 0$, where $\rm{\hat{I}}$ is the $2\times2$ identity matrix and $H$ is the Hamiltonian defined by Eq.~\Hamiltonian in the main text.
Complex roots $k_{r,p}$ of the characteristic polynomial
\begin{equation}
 \det\left[H - \rm{\hat{I}}E\right] = \sum\limits_{k=0}^6a(E, q)k_{r}^k\;\label{eq:polynom}
\end{equation}
are equal to eigenvalues of the companion matrix \cite{companion} of this polynomial,
hence we find the roots $k_{r,p}$ by numerically diagonalizing the companion
matrix.
Then, the spinor component $\chi$ of the corresponding plane wave
$\Phi_{k_{r,p},q}(x,y) = e^{iqy}e^{ik_{r,p}x} \chi(k_{r,p},q)$ can be  numerically computed
from the eigenvalues problem
\begin{equation}
 H(k_{r,p})\chi(k_{r,p},q) = E\chi(k_{r,p},q) \label{eq:eigvec}
\end{equation}
for $p=1,\dots,6$.

\section{Fourier analysis of $\delta\rho(x)$}
\label{sec:fourier}

\begin{figure}
 \includegraphics[width=0.48\textwidth]{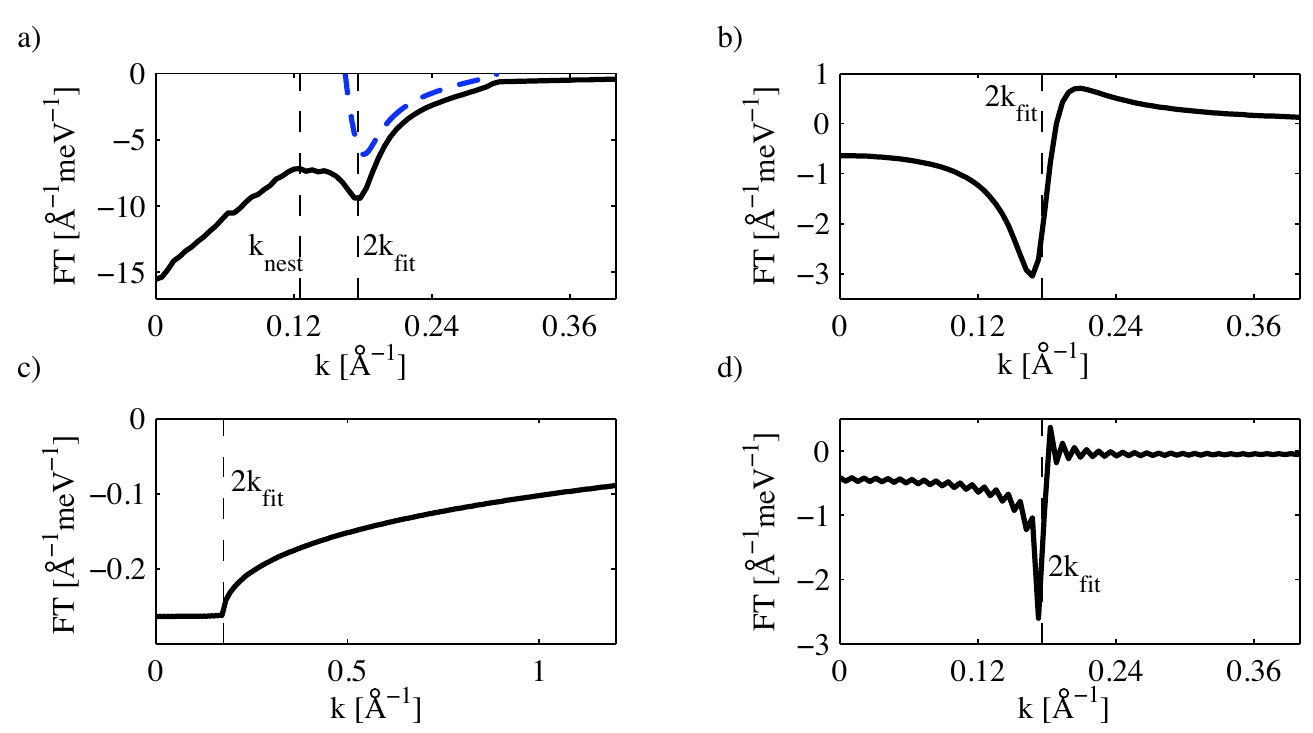}
  \caption{Comparison of the discrete Fourier transforms of 
(a) the computed LDOS oscillation $\delta\rho(x)$ shown in 
Fig. \ref{fig:results}a,
(b) the function $f_2(x)  = A_2 \sin(2k_{\rm fit} x + \varphi_2) e^{-x/L}$, where 
$A_2$, $k_{\rm fit}$, $\varphi_2$, $L$ are obtained from fitting $f_2$
to $\delta \rho$, 
(c) $A_2 \sin(2k_{\rm fit} x + \varphi_2) x^{-3/2}$, and
(d) $A_2 \sin(2 k_{\rm fit} x + \varphi_2) x^{-1/2}$.
} \label{fig:FFT}
\end{figure}

In Sec. \ref{sec:results},
we present numerical results for the defect-induced
spatial modulation of the LDOS, $\delta \rho(x)$. 
In order to develop an understanding of the decay characteristic
of the LDOS oscillations, we fit a power-law decaying 
($\delta \rho \propto x^{-3/2}$) as 
well as an exponentially decaying
function, $f_1$ and $f_2$, respectively, to our data.
According to Fig. \ref{fig:results}a, the  exponentially 
decaying $f_2$ provides a better fit, hinting that the decay
characteristics is closer to an exponential than to a power-law
predicted earlier\cite{LDOS_Zhou,LDOS_Biswas,Biswas_arxiv,
JingWang}.
However, our fitting procedure is not conclusive, 
as the exponentially decaying function $f_2$ 
has an extra fitting parameter, the length scale $L$ of the decay.

In order to investigate the decay characteristics
further, here
we provide the discrete Fourier transform (FT) 
(Fig. \ref{fig:FFT}a)
of the real-space data in Fig. \ref{fig:results}a,
and compare that to discrete Fourier transformed oscillations
that decay in an exponential (Fig. \ref{fig:FFT}b)
or power-law (Fig \ref{fig:FFT}c,d) fashion.
The discrete
Fourier transformation is carried out 
after symmetrization of the real-space data, i.e., after
mapping the real-space data set $f(x_i)$ $(i=0,\dots,N-1)$ to
$f_{{\rm sym},j}$ ($j = 0,\dots, 2N-1$) via the definition
\begin{equation}
f_{{\rm sym},j} :=
\left\{
\begin{array}{cl}
f(x_{j}) & \mbox{if $0 \leq j \leq N-1$,}
\\
f(x_{2N-1-j}) & \mbox{if $N \leq j\leq 2N-1$}. 
\end{array} \right.
\end{equation}
This symmetrization ensures that the Fourier transform will
be real valued in the large $N$ limit.

The subplots of Fig. \ref{fig:FFT} show the 
FT of
(a) our numerical results $\delta\rho(x)$ shown in 
Fig. \ref{fig:results}a,
(b) the function $f_2(x)  = A_2 \sin(2k_{\rm fit} x + \varphi_2) e^{-x/L}$, where 
$A_2$, $k_{\rm fit}$, $\varphi_2$, $L$ are obtained from fitting $f_2$
to $\delta \rho$, 
(c) $A_2 \sin(2k_{\rm fit} x + \varphi_2) x^{-3/2}$, and
(d) $A_2 \sin(2 k_{\rm fit} x + \varphi_2) x^{-1/2}$.
The data in Fig. \ref{fig:FFT}c,d shows, in accordance with the
known analytical formula describing the Fourier transform
of power-law decaying sinusoidal 
oscillations\cite{LDOS_Biswas},
that the FT develops 
non-analytical behavior at the characteristic wave number
$2 k_{\rm fit}$.
In contrast, our data set in Fig. \ref{fig:FFT}a
shows no such non-analytical behavior,
similarly to the FT 
of the exponentially decaying oscillation in Fig. \ref{fig:FFT}b.
This observation, although still not conclusive,
further supports the possibility that the LDOS oscillations
in the vicinity of the line defect do not follow a power-law
decay, and perhaps are closer to an exponential.
Further analytical studies, e.g., the extension of the asymptotic
analysis \cite{LDOS_Biswas,Biswas_arxiv,JingWang} 
to the pre-asymptotic spatial region might
help settling this open issue. 

In Sec.~\ref{sec:results}, 
we argued that the pre-asymptotic component of the 
LDOS is due to the interference of incoming and reflected partial waves
i.e., that the role of evanescent and transmitted partial waves is 
negligible.
To strengthen that point further, we plot the quantity 
\begin{equation}
\tilde \rho_r (2k) = 
\frac{{\rm Re}\left(r_{k,q'} \chi^\dag_{k,q'} \chi_{-k,q'}\right)}
{2 \pi \hbar |v_\parallel(E,k)|}
\end{equation}
as a function of $k$ [for definitions, see around Eq. \eqref{eq:rhor}],
as a dashed blue line in Fig.~\ref{fig:FFT}a.
Note that $\tilde \rho_r (2k)$ is
the Fourier transform of  the symmetrized 
$\rho_r(x)$ of Eq. \eqref{eq:rhor}.
Figure \ref{fig:FFT}a further supports the interpretation that 
the dip around the characteristic wave number
$2k_{\rm fit}$ of the LDOS oscillation forms as a result of interference of 
incoming and reflected partial waves.


\bibliographystyle{prsty}

\end{document}